\documentclass[twocolumn]{aastex631}
\usepackage{soul}
\usepackage{comment}
\usepackage{amsmath}
\usepackage{amssymb}
\usepackage{bbold}
\usepackage[super]{nth}

\DeclareMathOperator{\sign}{sign}

\begin{document}

\title{Fully Kinetic Simulations of Proton-Beam-Driven Instabilities from Parker Solar Probe Observations}

\author[0000-0001-5079-7941]{L. Pezzini}
\affiliation{Solar-Terrestrial Centre of Excellence--SIDC, Royal Observatory of Belgium, Brussels, Belgium}
\affiliation{Centre for mathematical Plasma Astrophysics, Department of Mathematics, KU Leuven, Leuven, Belgium}

\author[0000-0002-2542-9810]{A. N. Zhukov}
\affiliation{Solar-Terrestrial Centre of Excellence--SIDC, Royal Observatory of Belgium, Brussels, Belgium}
\affiliation{Skobeltsyn Institute of Nuclear Physics, Moscow State University, Moscow, Russia}

\author[0000-0002-7526-8154]{F. Bacchini}
\affiliation{Centre for mathematical Plasma Astrophysics, Department of Mathematics, KU Leuven, Leuven, Belgium}
\affiliation{Royal Belgian Institute for Space Aeronomy, Solar-Terrestrial Centre of Excellence, Brussels, Belgium}

\author[0000-0001-7233-2555]{G. Arrò}
\affiliation{Los Alamos National Laboratory, Los Alamos, NM 87545, USA}

\author[0000-0003-3223-1498]{R. A. López}
\affiliation{Research Center in the intersection of Plasma Physics, Matter, and Complexity ($P^2 mc$),\\ Comisi\'on Chilena de Energ\'{\i}a Nuclear, Casilla 188-D, Santiago, Chile}
\affiliation{Departamento de Ciencias F\'{\i}sicas, Facultad de Ciencias Exactas, Universidad Andres Bello, Sazi\'e 2212, Santiago 8370136, Chile}

\author[0000-0001-9293-174X]{A. Micera}
\affiliation{Institut f\"{u}r Theoretische Physik, Ruhr-Universität Bochum, Bochum, Germany}

\author[0000-0002-5782-0013]{M. E. Innocenti}
\affiliation{Institut f\"{u}r Theoretische Physik, Ruhr-Universität Bochum, Bochum, Germany}

\author[0000-0002-3123-4024]{G. Lapenta}
\affiliation{Centre for mathematical Plasma Astrophysics, Department of Mathematics, KU Leuven, Leuven, Belgium}
 
\begin{abstract}

The expanding solar wind plasma ubiquitously exhibits anisotropic non-thermal particle velocity distributions. Typically, proton Velocity Distribution Functions (VDFs) show the presence of a core and a field-aligned beam. Novel observations made by Parker Solar Probe (PSP) in the innermost heliosphere have revealed new complex features in the proton VDFs, namely anisotropic beams that sometimes experience perpendicular diffusion. In this study, we use a 2.5D fully kinetic simulation to investigate the stability of proton VDFs with anisotropic beams observed by PSP. Our setup consists of a core and an anisotropic beam populations that drift with respect to each other. This configuration triggers a proton-beam instability from which nearly parallel fast magnetosonic modes develop. Our results demonstrate that before this instability reaches saturation, the waves resonantly interact with the beam protons, causing perpendicular heating at the expense of the parallel temperature.

\end{abstract}


\section{Introduction} \label{sec:intro}
For decades, spacecraft missions like Helios, Ulysses and Wind have conducted in situ observations, measuring velocity distribution functions (VDFs) of both ions and electrons in the solar wind \citep{marschSolarWindProtons1982, marschKineticPhysicsSolar2006}. These observations revealed several features regarding the kinetic properties of the different particle species in the solar wind plasma. 
In this environment, not all plasma species have equal temperatures. Electrons are generally colder than protons in the fast solar wind but hotter than protons in the slow solar wind \citep{montgomerySolarWindElectrons1968, newburyElectronTemperatureAmbient1998}.
Proton and electron populations can also present temperature anisotropy with respect to the direction of the local magnetic field \citep{baleELECTRONHEATCONDUCTION2013}. Temperature anisotropies in the solar wind can be generated by large-scale mechanisms such as solar wind expansion \citep{phillipsRadialEvolutionSolar1990, stverakElectronTemperatureAnisotropy2008, opieConditionsProtonTemperature2022}, as well as by local wave-particle interactions \citep{tsurutaniBasicConceptsWaveparticle1997, verscharenINSTABILITIESDRIVENDRIFT2013, verscharenSelfinducedScatteringStrahl2019}. According to the double-adiabatic expansion model, in the absence of heat flux and collisions, protons in the inner heliosphere would show $T_{\perp, \mathrm{p}} < T_{\parallel, \mathrm{p}}$ \citep{chewBoltzmannEquationOnefluid1956}. Here, $T_{\perp, \mathrm{p}}$ and $T_{\parallel, \mathrm{p}}$ denote the perpendicular and parallel temperatures of protons, respectively. However, observations reveal a predominance of cases where $T_{\perp, \mathrm{p}} > T_{\parallel, \mathrm{p}}$ in the fast solar wind within the inner heliosphere, underscoring the need for a mechanism driving substantial perpendicular heating \citep{richardsonRadialEvolutionSolar1995, matteiniEvolutionSolarWind2007}. In \citealt{hellingerHeatingCoolingProtons2011, hellingerProtonThermalEnergetics2013} and \citealt{matteiniSignaturesKineticInstabilities2013}, the radial dependence of proton moments (e.g., density, bulk velocity, etc.) from Helios observations is fitted into the fluid equations for the parallel and perpendicular proton temperatures. These fits are used to calculate the heating (or cooling) rates in the parallel and perpendicular directions required to explain the observed radial trends in both the slow and the fast solar wind. That study finds that protons generally need to be heated perpendicularly between 0.3 and 1 AU, and that the necessary perpendicular proton heating rate is consistent with turbulent heating, as shown in \citealt{naritaGaryPictureShortWavelength2022, bowenMediationCollisionlessTurbulent2024}.
Proton VDFs are predominantly gyrotropic, often exhibiting skewness, resulting in a heat flux aligned along the magnetic field, directed away from the Sun \citep{marschSolarWindProtons1982}.
They also often exhibit a magnetic field-aligned beam in the anti-sunward direction, a secondary proton population streaming at higher velocities compared to the core one \citep{asbridgeAbundanceDifferencesSolar1974, altermanComparisonAlphaParticle2018}. It has been suggested that the process at the origin of the proton beam is the reflection of turbulent fluctuations at the MHD/kinetic spectral break: the greater the wavenumber of the spectral break, the faster the generated proton beam \citep{voitenkoGenerationProtonBeams2015}. These non-Maxwellian features are more pronounced in the fast solar wind compared to the slow solar wind, and their magnitude increases as the distance to the Sun decreases \citep{verscharenMultiscaleNatureSolar2019}. They represent symmetry-breaking properties capable of inducing plasma instabilities, leading to subsequent energy transfer processes restoring the equilibrium within the system \citep{vernieroParkerSolarProbe2020}.

The Parker Solar Probe (PSP) mission \citep{foxSolarProbeMission2016} aims to study fundamental processes involved in energy transfer within the solar wind and the solar Corona, such as acceleration and heating processes, to the solar surface. Observations from the Solar Wind Electron Alpha and Proton (SWEAP) instrument onboard PSP \citep{kasperSolarWindElectrons2016} detected anisotropic beams in the proton VDFs, as reported by \citealt{kleinInferredLinearStability2021}. Moreover, extreme cases of anisotropic beam structures were also observed, specifically termed ``hammerhead'' \citep{vernieroStrongPerpendicularVelocityspace2022}.
Consistently with the measurements of SWEAP, the FIELDS instrument \citep{baleFIELDSInstrumentSuite2016} detected a strong wave activity from Magnetohydrodynamics (MHD) scales down to ion kinetic scales \citep{bowenIonscaleElectromagneticWaves2020, vernieroParkerSolarProbe2020}. The solar wind is a weakly collisional plasma \citep{altermanComparisonAlphaParticle2018, verscharenMultiscaleNatureSolar2019}, thus it is wave/particle interaction processes, rather than collisions, which act in the direction of reducing the non-Maxwellian character of VDFs \citep{kasperWindSWEObservations2002, kleinMajoritySolarWind2018}. The linear Alfv\'en dispersion relation at MHD scales connects with the fast-magnetosonic/whistler (FM/W) waves at kinetic scales \citep{belcherLargeamplitudeAlfvenWaves1971, verscharenMultiscaleNatureSolar2019}. Linear theory shows that, in the direction parallel to the background magnetic field ($\boldsymbol{k} \times \boldsymbol{B}_{0}=\boldsymbol{0}$), three different instabilities can arise in the presence of an ion core and a less dense ion beam population: the left-hand (LH) resonant, the non-resonant, and the right-hand (RH) resonant ion–ion instabilities \citep{garyTheorySpacePlasma1993}. The observed relative drift speeds between core protons, secondary proton beams, and $\alpha$-particles \citep{marschSolarWindProtons1982, neugebauerUlyssesObservationsDifferential1996, steinbergDifferentialFlowSolar1996, podestaEFFECTDIFFERENTIALFLOW2011, podestaMAGNETICHELICITYSPECTRUM2011} constitute another reservoir of free energy that can excite waves through wave-particle resonant interaction or cyclotron resonance \citep{bourouaineLIMITSALPHAPARTICLE2013, verscharenINSTABILITIESDRIVENDRIFT2013, ofmanModelingIonBeams2022, ofmanObservationsModelingUnstable2023}. Indeed, proton beams, usually guided by the local magnetic field, have been observed in association with enhanced low-frequency electromagnetic fluctuations \citep{wicksPROTONCYCLOTRONWAVESTORM2016, kleinLinearStabilityInner2019, martinovicIondrivenInstabilitiesInner2021}. These observations suggest that the fluctuations are produced by ion beam instabilities \citep{verscharenMultiscaleNatureSolar2019, zhuNonfieldalignedProtonBeams2023}.
Damping mechanisms such as cyclotron resonant damping \citep{goldsteinPropertiesFluctuatingMagnetic1994, leamonContributionCyclotronresonantDamping1998} dissipate energy at ion kinetic scales, where particles interact efficiently with electromagnetic waves \citep{leamonContributionCyclotronresonantDamping1998, chandranINCORPORATINGKINETICPHYSICS2011}. Through quasi-linear theory it is possible to detect signatures of cyclotron resonant damping in proton density phase-space \citep{hollwegGenerationFastSolar2002, cranmerENSEMBLESIMULATIONSPROTON2014, bowenSituSignatureCyclotron2022}.

The highly anisotropic beam structure analyzed by \citealt{vernieroStrongPerpendicularVelocityspace2022}, referred to as the ``hammerhead’’, is believed to originate from resonant interaction between parallel-propagating fluctuations and protons at kinetic scales, similar to those observed in \citealt{kleinInferredLinearStability2021}. However, its precise origin has yet to be firmly established. Hybrid numerical simulations can efficiently study non-linear processes associated with unstable proton-beam VDFs \citep{ofmanModelingIonBeams2022, ofmanObservationsModelingUnstable2023}. However, in hybrid codes, electrons are treated as a fluid, which leads to a loss of their fundamental kinetic nature. We used a fully kinetic approach in which both protons and electrons are treated kinetically. This approach goes beyond the state of the art, where usually hybrid simulations are performed and provides a first verification of whether electron dynamics, in the specific physical scenario we consider, influences the issue of perpendicular proton heating in solar wind plasma. To achieve this, we extend and complement the stability analysis of the proton-beam unstable VDF investigated by \citealt{kleinInferredLinearStability2021} by exploring its nonlinear regime. In this study, we employ \textsc{ECsim} \citep{lapentaExactlyEnergyConserving2017, gonzalez-herreroECsimCYLEnergyConserving2019, bacchiniRelSIMRelativisticSemiimplicit2023, croonenExactlyEnergyconservingElectromagnetic2024}, a semi-implicit, exactly energy-conserving Particle-in-Cell (PIC) simulation tool, to investigate how the dynamics of both electrons and protons impact the development of the anisotropic proton beam. This paper is organized as follows: in Section~\ref{sec:setup}, we present the setup of the PIC simulation in detail. Section~\ref{sec:results} showcases the numerical modeling results alongside linear-theory analyses. Finally, Section~\ref{sec:discussion} contains the discussion and conclusions.

\section{Numerical setup} \label{sec:setup}

The \textsc{ECsim} code \citep{lapentaExactlyEnergyConserving2017,gonzalez-herreroECsimCYLEnergyConserving2019,bacchiniRelSIMRelativisticSemiimplicit2023,croonenExactlyEnergyconservingElectromagnetic2024} was employed to solve the Vlasov-Maxwell system of equations for a proton-electron plasma. Our initial parameters are inspired by PSP measurements made on the January \nth{30}  2020, as reported by \citealt{kleinInferredLinearStability2021}. We conducted a simulation in a periodic two-dimensional $(x, y)$ Cartesian geometry. At initialization, the electric field is set to zero and the background magnetic field is homogeneous, $\boldsymbol{B}_{0}= B_0 \hat{\boldsymbol{e}}_{x}$. Introducing a field-aligned reference system, the $x$-axis denotes the parallel direction, while the $y$- and $z$-axes represent the perpendicular directions.

In our setup, the protons, denoted by the subscript ``p'', have constant number density $n_{\mathrm{p}}$ and are divided into two sub-populations. The core, with subscript ``c'', has number density $n_{\mathrm{c}}/n_{\mathrm{p}}=0.864$, while the beam, subscript ``b'',  has number density $n_{\mathrm{b}}/n_{\mathrm{p}}=0.136$. The core and the beam are both initialized with a bi-Maxwellian distribution, each characterized by a non-zero drift velocity along the $x$-axis (parallel direction):

 \begin{equation}
        \begin{aligned}
            f_{\mathrm{p}}\left(v_{x}, v_{y}, v_{\mathrm{z}}\right) = & \sum_{i=\mathrm{c}, \mathrm{b}} \frac{n_i}{\pi^{3 / 2}w_{x,i} w_{y,i} w_{z,i}} \\
            & \times \exp \left(-\frac{(v_{x}-V_{x, i})^2}{w_{x,i}^2}-\frac{v_{y}^2}{w_{y,i}^2}-\frac{v_z^2}{ w_{z,i}^2} \right).
        \end{aligned}
        \label{eq:maxwell}
\end{equation}

In equation~\eqref{eq:maxwell}, the subscript $i$ denotes the proton species, i.e.\ the core or the beam. The thermal velocities of the $i$-th proton species are expressed as $w_{i}\doteq \sqrt{k_{\mathrm{B}}T_{i}/m_{\mathrm{p}}}$, where $m_{\mathrm{p}}$ stands for the proton mass. Initially, these velocities are set to $w_{x, \mathrm{c}}/c_{\mathrm{A, p}} \approx 0.64$ and $w_{y, \mathrm{c}}/c_{\mathrm{A, p}} \approx 0.56$ for the core, while for the beam they are chosen as $w_{x, \mathrm{b}}/c_{\mathrm{A, p}}\approx 1.00$ and $w_{y, \mathrm{b}}/c_{\mathrm{A, p}}\approx0.79$. The Alfv\'en speed for the protons is defined as $c_{\mathrm{A,p}}\doteq B_{0}/\sqrt{4 \pi n_{\mathrm{p}}m_{\mathrm{p}}}$ at initialisation with a value of $c_{\mathrm{A,p}}/c=0.00291$, where $c$ represents the speed of light in vacuum. This results in the following temperature anisotropies: $T_{\perp, \mathrm{c}}/T_{\parallel, \mathrm{c}} = 0.770$ and $T_{\perp, \mathrm{b}}/T_{\parallel, \mathrm{b}} = 0.620$. Given that the plasma is described by a bi-Maxwellian distribution, there exists a symmetry in the perpendicular velocities, expressed as $w_{y, i} = w_{z, i}$ for all species $i$. The drift velocities of the core and beam are approximately $V_{x, \mathrm{c}}/c_{\mathrm{A, p}} \approx 0.18$ and $V_{x, \mathrm{b}}/c_{\mathrm{A, p}} \approx - 1.16$, respectively. Electrons are initialized as a Maxwellian with thermal velocity $w_{\mathrm{e}}/c_{\mathrm{A, p}} \approx 7.12$ with unitary number density $n_{\mathrm{e}}$. The described electron-proton plasma satisfies the zero-net-current condition $n_{\mathrm{c}} V_{x, \mathrm{c}} + n_{\mathrm{b}} V_{x, \mathrm{b}} = 0$ and the quasi-neutrality condition $n_{\mathrm{e}}\approx n_{\mathrm{p}} = n_{\mathrm{c}} + n_{\mathrm{b}}$.

The proton VDF previously discussed is unstable and expected to produce fast magnetosonic modes, as demonstrated by the linear analysis conducted in \citealt{kleinInferredLinearStability2021}. Based on linear theory calculations (for a detailed analysis, see Section~\ref{sec:results}), the instability would manifest within a specific range of spatio-temporal scales, namely around those associated with the fastest growing mode. Consequently, we adjust the spatial and temporal parameters of our simulation accordingly. The fastest growing mode has a parallel wavenumber $k_{\parallel}/d_{\mathrm{p}}^{-1}\approx0.501$ and a perpendicular one $k_{\perp}/d_{\mathrm{p}}^{-1}\approx0.0245$. Here, $d_{\mathrm{p}}$ is the proton skin depth, defined as $d_{\mathrm{p}} \doteq c/\omega_{\mathrm{p}}$, and $\omega_{\mathrm{p}}\doteq \sqrt{4 \pi e^2 n_{\mathrm{p}} / m_{\mathrm{p}}}$ is the proton plasma frequency. This leads to a wavelength in the parallel direction $\lambda_{x}/d_{\mathrm{p}} = 2 \pi/(k_{\parallel}/d_{\mathrm{p}}^{-1})\approx 13$ and in the perpendicular direction $\lambda_{y}/d_{\mathrm{p}} = 2 \pi/(k_{\perp}/d_{\mathrm{p}}^{-1})\approx 209$. Therefore, the domain length in each direction is selected to fit several wavelengths of this mode. This choice yields a rectangular box with dimensions $L_{x}/d_{\mathrm{p}}=64$ and $L_{y}/d_{\mathrm{p}}=256$. The numerical grid consists of $256 \times 1024$ cells with a spatial increment of $\Delta x/d_{\mathrm{p}}=\Delta y/d_{\mathrm{p}}=0.25$. Similarly to the spatial domain, the estimation of the temporal scale is made by employing the growth rate of the most unstable mode $\gamma_{\mathrm{m}}/ \omega_{\mathrm{p}} \approx 0.000121$ or $\gamma_{\mathrm{m}}/ \Omega_{\mathrm{p}} \approx 0.042$, where $\Omega_{\mathrm{p}} \doteq e B_0 / (m_{\mathrm{p}} c)$ is the proton cyclotron frequency. The instability evolves over a time scale $t/\Omega_{\mathrm{p}}^{-1} = 2 \pi/(\gamma_{\mathrm{m}}/\Omega_{\mathrm{p}}) \approx 151.15$. Therefore, the simulation is run for a total time $t_{\mathrm{tot}}/\omega_{\mathrm{p}}^{-1}\approx 436$, to fully capture the development of the instability until the non-linear phase. The time step is set to $\Delta t /\omega_{\mathrm{p}}^{-1} = 0.00145$ to ensure accurate resolution of the proton gyro-period. We employed $4096$ particles per cell per species, initially distributed uniformly on the grid. We utilized a reduced proton-to-electron mass ratio, $m_{\mathrm{p}}/m_{\mathrm{e}} = 183.6$, to narrow the gap between electron and proton dynamical scales, thereby saving computational time.

\section{Results}\label{sec:results}

\begin{figure*}
\includegraphics[width=1\textwidth]{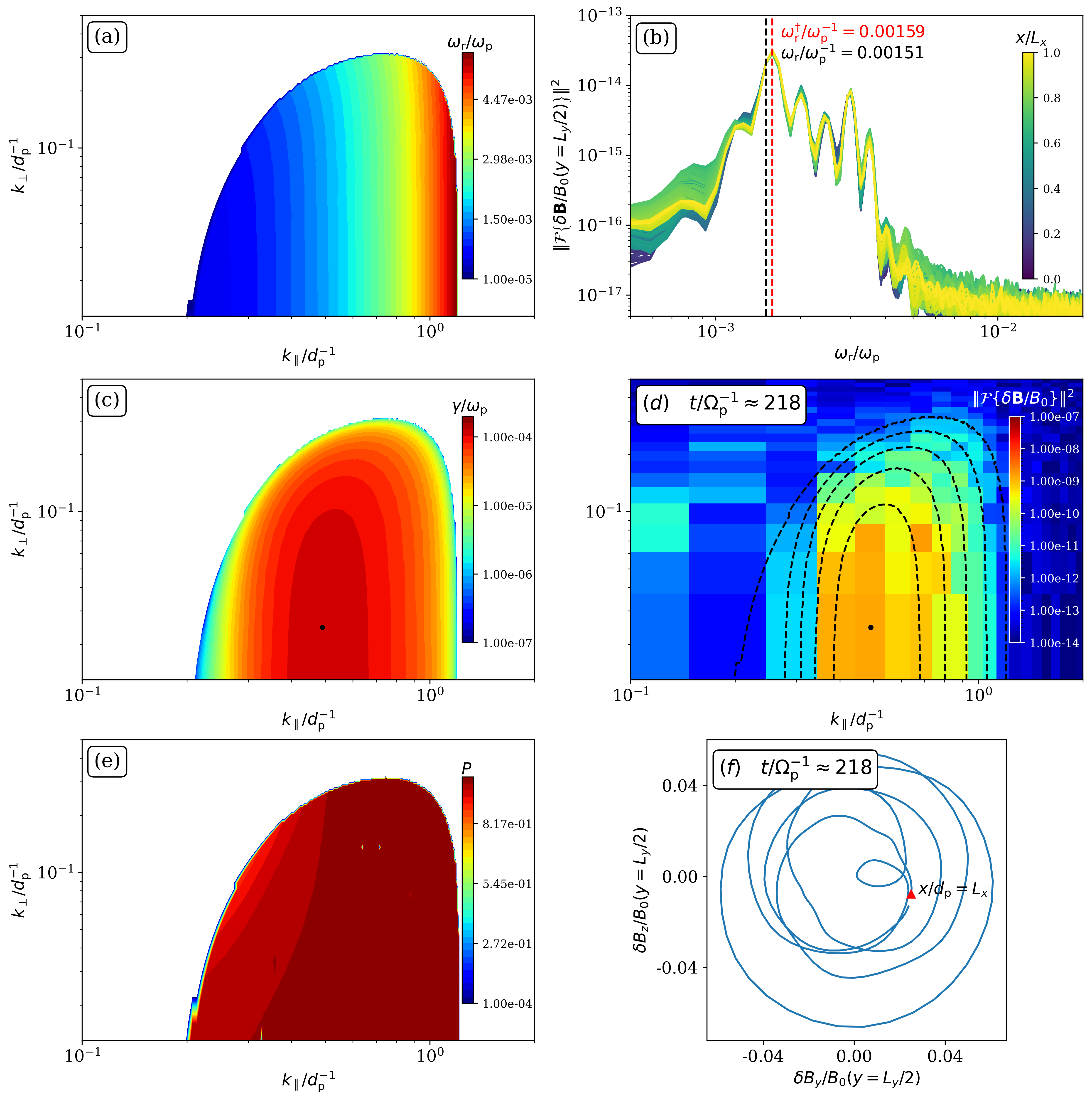}
\caption{Results from linear Vlasov theory corresponding to our initial setup, in the left column, compared with the simulation results in the right column. Panels (a) to (f) show the following: the real frequency $\omega_{\mathrm{r}}/\omega_{\mathrm{p}}$ (a); the imaginary frequency $\gamma$ (c); the polarization of the wave (e). The right column shows the following: in panel (b), the power spectrum measured for a cut at $y=L_{y}/2$ is plotted against the frequency at different positions along the $x$-axis, the black and red dashed lines represent the frequencies obtained from the linear theory and the numerical simulation, respectively; panel (d) shows the power spectrum at $t/\Omega_{\mathrm{p}}^{-1}\approx 218$, with some of the instability growth rates contours of panel (c) overlaid, with the black dot denoting the fastest-growing mode; panel (f) is the hodogram at $t/\Omega_{\mathrm{p}}^{-1}\approx 218$, where the red triangle shows the value of the magnetic fluctuation corresponding to the point $x=L_{x}$ (see text for details on how this is computed).}
\label{fig:modes}
\end{figure*}

The results from our numerical simulation are compared with Vlasov linear theory in Figure~\ref{fig:modes}. We solved the linear dispersion relation using the DIS-K solver\footnote{The code is publicly available and can be found at \url{https://github.com/ralopezh/dis-k}.} for the setup described above \citep{lopezGeneralDispersionProperties2021,lopezRalopezhDiskFirst2023}. Specifically, the spectrum and polarization of the waves developed from our initial setup are analyzed. The real part $\omega_{\mathrm{r}}/\omega_{\mathrm{p}}$ of the complex frequency, as obtained from linear theory, is shown in Figure~\ref{fig:modes}(a) against $k_{\parallel}$ and $k_{\perp}$. In Figure~\ref{fig:modes}(b) we plot the power spectrum of magnetic fluctuations $\delta\boldsymbol{B}/B_{0}=(\boldsymbol{B}-\boldsymbol{B}_{0})/B_{0}$, measured along the parallel direction fixing $y=L_{y}/2$, denoted as $\delta\boldsymbol{B}(y=L_{y}/2)/B_{0}$. Here, $\mathcal{F}$ represent the Fast Fourier Transform (FFT) in time, and the colorbar indicates the variation in the spatial dimension $x/L_{x}$. In our simulation, the peak of maximum intensity occurs at $\omega^{\dagger}_{\mathrm{r}}/\omega_{\mathrm{p}} = 0.00159$ in red, while the value calculated via quasi-linear theory is  $\omega_{\mathrm{r}}/\omega_{\mathrm{p}} = 0.00151$ in black, as pointed out in panel (b) by the black dashed line, corresponding to $k_{\parallel}/d_{\mathrm{p}}^{-1} \approx 0.501$ in panel (a). Comparisons between linear theory and simulation results reveal small discrepancies, as the VDF remains constant in linear theory but is deformed as it evolves during the course of the simulation.

\begin{figure*}
\includegraphics[width=1\textwidth]{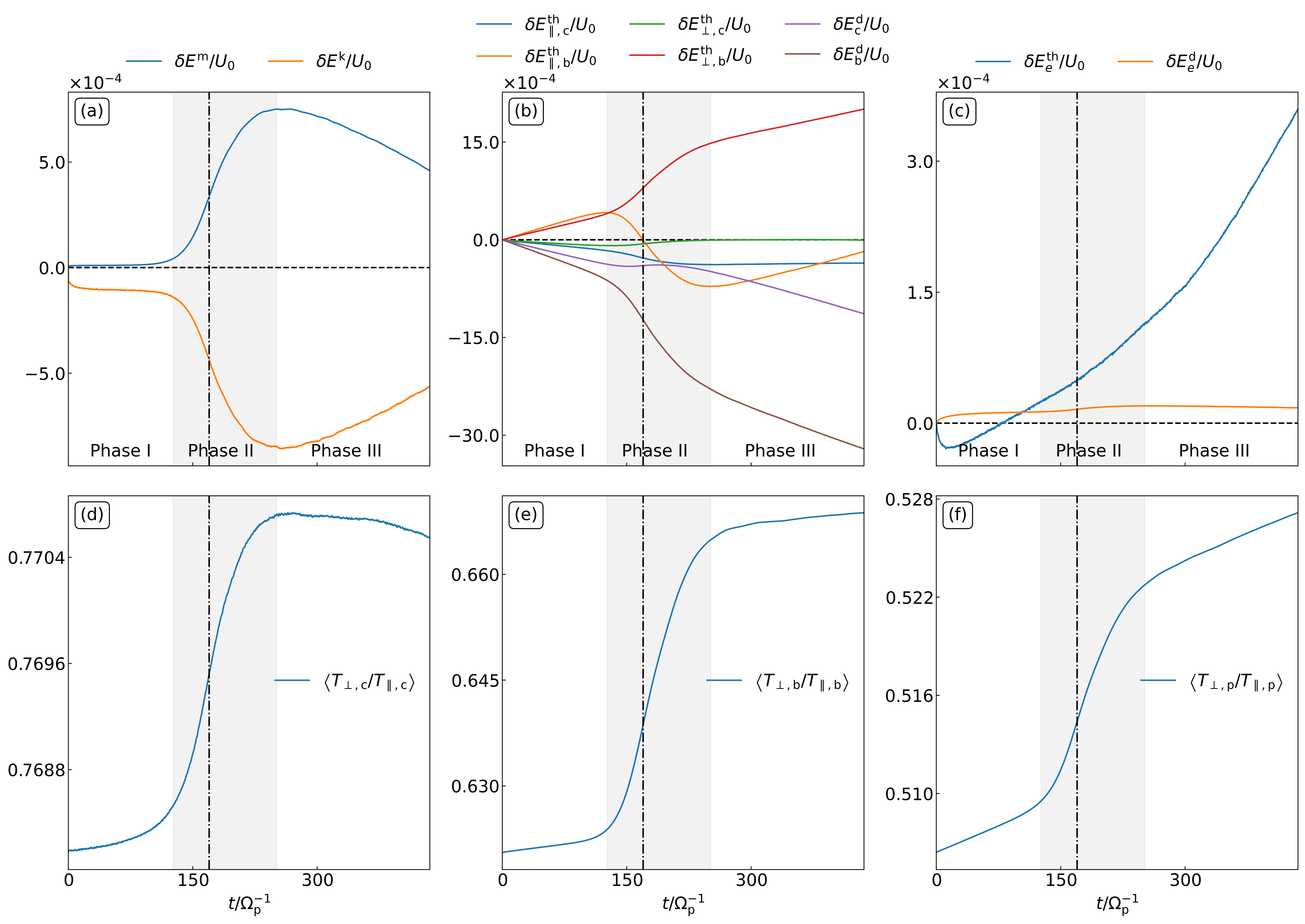}
\caption{Evolution of different types of energy and  proton temperature anisotropy. Panels (a)-(c) illustrate the time evolution of different energy components. In panels (d)-(f), the temporal evolution of the temperature anisotropy for the proton core, proton beam, and the total proton population (encompassing both core and beam) is displayed. The three distinct phases of the instability—labeled I, II, and III—corresponding to the stationary, linear, and non-linear phases, are also depicted. The gray region identifies phase II, with the dashed-dotted line indicating the inflection point of the curves, where the conversion of kinetic energy into magnetic energy peaks at $t_{\mathrm{peak}}/\Omega_{\mathrm{p}}^{-1}\approx 170$. Panel (a) displays the changes in magnetic and total kinetic energy of electrons and protons, normalized by the initial total energy of the system. Panel (b) presents the thermal energies (parallel and perpendicular with respect to the guide field direction) of both the core and beam populations, along with the core and beam drift (i.e.\ bulk) energy. In panel (c), the thermal and bulk energies of the electron population are shown.} The formal definition of these quantities is provided in the text.\label{fig:energy}
\end{figure*}

Figure~\ref{fig:modes}(c) shows the power spectrum of $\gamma/\omega_{\mathrm{p}}$ in the $k_{\parallel}$--$k_{\perp}$ plane. The dominant mode (i.e.\ fastest-growing mode), indicated by the black dot, propagates mostly along the parallel direction ($x$-axis) with $k_{\parallel}/d_{\mathrm{p}}^{-1} \approx 0.501$ and $k_{\perp}/d_{\mathrm{p}}^{-1} \approx 0.0245$, having a magnitude of $\gamma_{\mathrm{m}}/\Omega_{\mathrm{p}}\approx0.042$. In Figure~\ref{fig:modes}(d), we present the analog of panel (c) for the simulated data. We display the power spectrum of $\delta\boldsymbol{B}/B_0$ at $t/\Omega_{\mathrm{p}}^{-1} \approx 218$ (end of the quasi-linear phase). Similarly to panel (b), $\mathcal{F}\left\{ \delta \boldsymbol{B}/B_0\right\}$ represents the 2D FFT in the spatial domain of $\delta\boldsymbol{B}/B_0$. Black isocontours of $\gamma/\omega_{\mathrm{p}}$ obtained from linear theory, as in panel (c), have been superimposed in panel (d) to highlight the good agreement between analytic and simulation results. This specific time has been chosen because it corresponds to the moment when the magnetic energy reaches its maximum, see Figure~\ref{fig:energy}(a). 

In Figure~\ref{fig:modes}(e), we plot the polarization $P \doteq \mathbb{R} \big\{ i {\rm \sign}(\omega_\mathrm{r})D_x/D_y \big\}$ \citep{stixWavesPlasmas1962,garyTheorySpacePlasma1993}, calculated from linear theory, where $D_{x}$ and $D_{y}$ are the electric field component of the wave. If $P>0$, then the wave is right-handed (RH) elliptically polarised, whereas if $P<0$, it is left-handed (LH). In this case, the polarization is RH, but it is very close to unity, making the expected hodogram almost circular. The hodogram in Figure~\ref{fig:modes}(f) provides information about the polarization of the wave for our simulation. By the reported time frame of $t/\Omega_{\mathrm{p}}^{-1}\approx 218$ all unstable modes have completed their development. A wave polarization state can be split into two linearly polarized orthogonal components with respect to the guiding magnetic field \citep{naritaReviewArticleWave2017,bornPrinciplesOptics2019}. In our case, the two components are measured as the values, along the $x$-axis at $y=L_{y}/2$, of $\delta B_{y}/B_{0}$ and $\delta B_{z}/B_{0}$, representing the magnetic fluctuation in the transverse directions at $t/\Omega_{\mathrm{p}}^{-1}\approx 218$. The red triangle highlights the value of the magnetic fluctuation at $x/{d_{\mathrm{p}} }=L_{x}$ and $y/{d_{\mathrm{p}} }= L_{y}/2$. From this point, a counterclockwise rotation corresponds to an RH circular polarization state, while a clockwise rotation implies an LH circular polarization state. In Figure~\ref{fig:modes}(f) we have an RH circularly polarized wave, which once again aligns with the linear theory shown in Figure~\ref{fig:modes}(e) and discussed in \citet{kleinInferredLinearStability2021}.

\begin{figure*}
\includegraphics[width=1\textwidth]{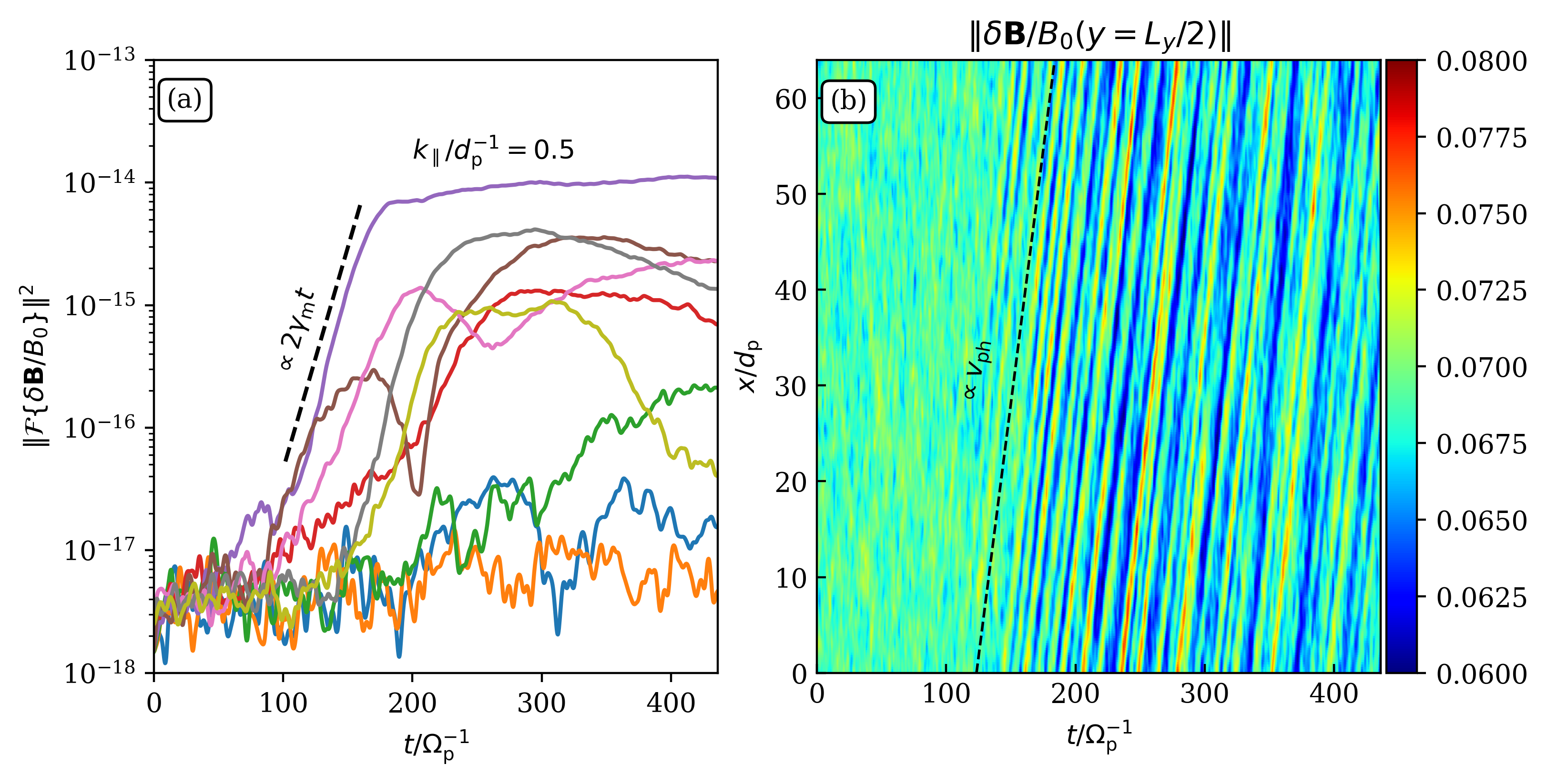}
\caption{Magnetic field fluctuations resulting from the PIC simulation. Panel (a) displays the spectral power of magnetic fluctuations for different wavenumbers (i.e., different modes), corresponding to those within the region delimited by the innermost contour in Figure~\ref{fig:modes}(d) with $k_{\parallel}/d_{\mathrm{p}}^{-1}\in [0.3, 0.6]$. The calculated fastest-growing mode, represented by the violet line and highlighted with its corresponding wavenumber, is compared with the expected maximum growth rate from linear theory, indicated by the dashed black line. In Panel (b), a spatio-temporal analysis is presented, focusing on the cut along $y=L_{y}/2$ of the magnetic fluctuation stacked in time. The black line in panel (b) has a slope equal to the phase velocity of the wave calculated from linear theory. \label{fig:fgm}}
\end{figure*}

\begin{figure*}
\includegraphics[width=1\textwidth]{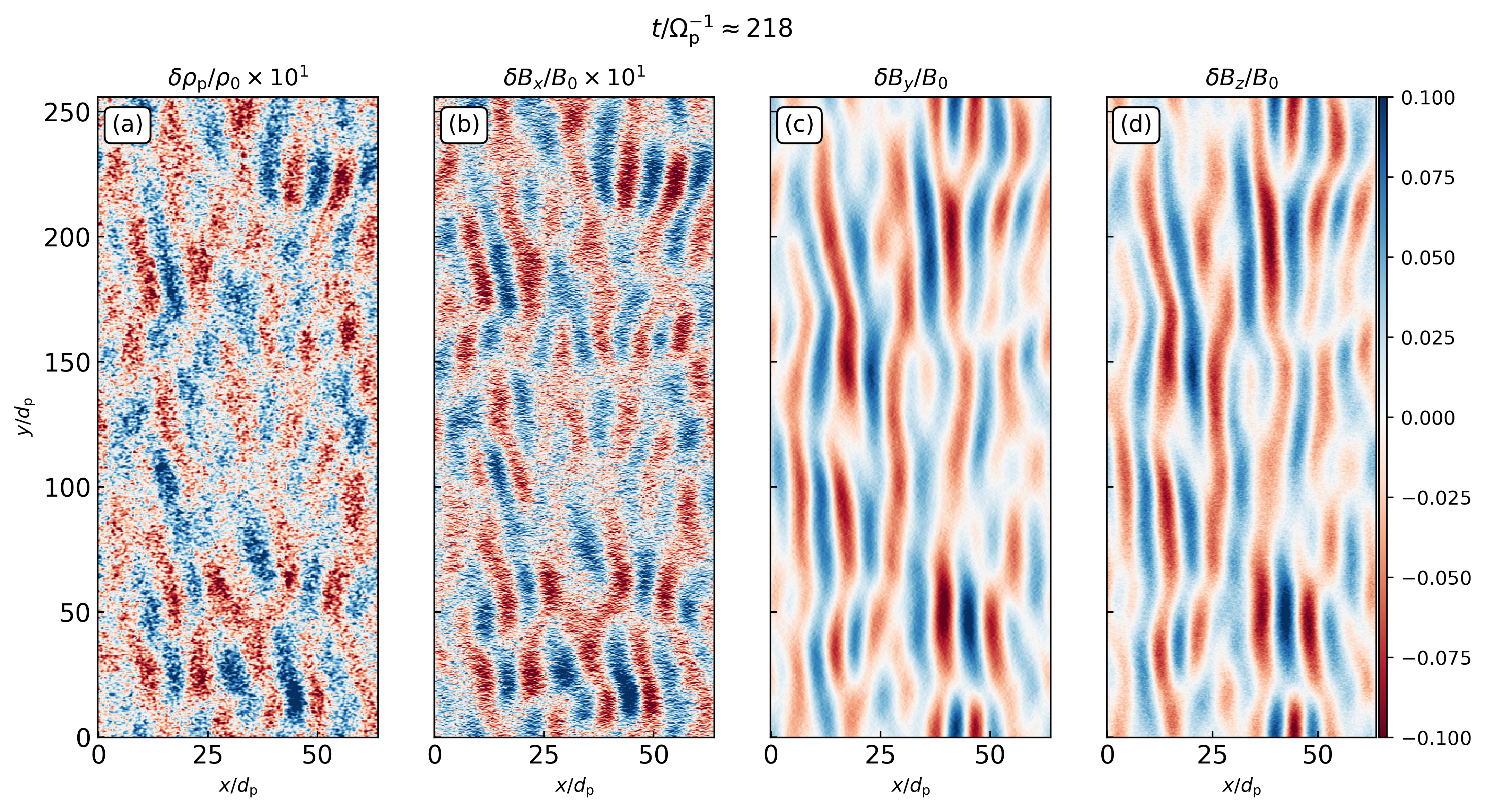}
\caption{Fluctuations in the proton density $\rho_{\mathrm{p}}$ in panel (a), with $\delta \rho_{\mathrm{p}}=\rho_{\mathrm{p}}(t)-\rho_{0}$, and fluctuations in the magnetic field components $B_{x}, B_{y}, B_{z}$, in panel (b)-(d), at the moment of instability saturation $t/\Omega_{\mathrm{p}}^{-1} \approx 218$. The amplitude of $\rho_\mathrm{p}$ and $B_x$ fluctuations have been multiplied by a factor 10 for visualization purposes.}
\label{fig:fields}
\end{figure*}

In Figure~\ref{fig:fgm}(a), the spectral power of magnetic fluctuations has been plotted for a number of unstable modes within the range of parallel wavenumbers $k_{\parallel}/d_{\mathrm{p}}^{-1}\in [0.3, 0.6]$ considered in Figure~\ref{fig:modes}(d). Our main focus is on the mode with the steepest growth, representing the fastest-growing one. This mode possesses wavenumber $k_{\parallel}/d_{\mathrm{p}}^{-1} = 0.501$ and $k_{\perp}/d_{\mathrm{p}}^{-1} = 0.0245$ along with the growth rate $\gamma_{\mathrm{m}}/\Omega_{\mathrm{p}}=0.042$, as determined by linear theory. The maximum theoretical growth rate $\gamma_{\mathrm{m}}$ from the linear analysis of Figure~\ref{fig:modes}(c) is shown as a dashed line for comparison. The fastest growing mode is exponentially growing during the quasi-linear phase of the instability, occurring from time $t/\Omega_{\mathrm{p}}^{-1} \approx 116$ to $t/\Omega_{\mathrm{p}}^{-1} \approx 189$, closely matching the predicted growth rate according to Vlasov linear theory. From the analysis in Figure~\ref{fig:modes}(b)-(c), we can calculate the phase velocity of the wave corresponding to the most-unstable mode, denoted as $v_{\mathrm{ph}}=\omega_{\mathrm{r}}/k_{\parallel}\approx 0.00306$. In Figure~\ref{fig:fgm}(b), is presented a spatial and temporal analysis $\delta \boldsymbol{B}/B_{0}(y=L_{y}/2)$. This plot was obtained by stacking in time the values of $\delta \boldsymbol{B}/B_{0}(y=L_{y}/2)$. The line we overlaid, with equation $x = v_{\mathrm{ph}}t + b$, where $b/d_\mathrm{p} \approx -130.6$ is the offset, demonstrates that the perturbation in the magnetic field is propagating at the speed predicted by the linear analysis.

The spatial distribution of fluctuations in proton mass density $\rho_\mathrm{p}$ and magnetic field components is displayed in Figure~\ref{fig:fields}(a)--(d). They are presented at time $t/\Omega_{\mathrm{p}}^{-1}\approx 218$, corresponding to the moment of instability saturation after the end of the quasi-linear phase, to examine the waves when they are fully developed. The perturbation in proton density, as visible in panel (a), indicates that the instability develops as a compressive wave. Additionally, there is a positive correlation between the fluctuation in proton density and the magnetic field fluctuation in the $x$-direction, as denoted by both the pattern and intensity shown in Figure~\ref{fig:fields}(a)-(b). The magnetic fluctuation in the $y$- and $z$-directions, representing the transverse directions, is more intense, indicating that it carries more energy than that in the parallel direction. We observe that these wave packets are predominantly parallel-propagating, and thus $\theta \doteq k_{\perp}/k_{\parallel} \approx 0.05$, with $\theta$ representing the angle between the wavevector components $k_{\parallel}$ and $k_{\perp}$ as calculated from linear theory. Therefore, we can conclude that the developed wave is a fast-magnetosonic/whistler (FM/W) wave.

In Figure~\ref{fig:energy}, we show the energy exchanges within the considered system. In Figure~\ref{fig:energy} (a)-(f) we identify three phases of evolution the system is passing through. Phase I or \textit{quasi-stationary phase} ends at approximately $t_{\mathrm{beg}}/\Omega_{\mathrm{p}}^{-1} \approx 126$, which coincides with the moment the instability starts growing with the fastest growing mode. Subsequently, Phase II or \textit{quasi-linear phase}, in light gray, begins at the end of phase I and evolves until the instability saturates. Then, the conversion from magnetic to kinetic energy reaches its maximum around $t_{\mathrm{end}}/\Omega_{\mathrm{p}}^{-1} \approx 251$ as illustrated in Figure~\ref{fig:fgm}(a). At the end of this phase, the total magnetic and kinetic energies respectively reach their maximum and minimum values. Moreover, the dash-dotted line in Figure~\ref{fig:energy} highlights the inflection point of the energy variation, which is the point where the conversion of kinetic energy into magnetic energy peaks at $t_{\mathrm{peak}}/\Omega_{\mathrm{p}}^{-1}\approx 170$. It is followed by Phase III, named \textit{non-linear phase}, which starts at $t_{\mathrm{end}}/\Omega_{\mathrm{p}}^{-1} \approx 251$ and continues until the end of the simulation.

In Figure~\ref{fig:energy}(a)-(c) we plot the difference in each energy $\delta E = E(t)-E(t = 0)$ normalized by $U_{0}$, which is the total energy of the system at initialization, using the formula $\delta E/U_{0} \doteq \left(E(t) - E(t_0)\right)/U_{0}$. The \textit{quasi-stationary phase} is characterized by a transient in which the total magnetic energy $E^{\mathrm{m}} \doteq \int B^2/(8 \pi) ~\mathrm{d}^3 \boldsymbol{x}$ and the total kinetic energy of electrons and protons $E^{\mathrm{k}} = \sum_{s=\mathrm{e},\mathrm{p}} \left( E^{\mathrm{th}}_{s} + E^{\mathrm{d}}_{s} \right)$ remain nearly constant over time, as shown in Figure~\ref{fig:energy}(a). The kinetic thermal energy for species $s$ is defined as $E^{\mathrm{th}}_{s} \doteq (1/2) \int \mathrm{Tr}(\boldsymbol{P}_{s}) ~\mathrm{d}^3 \boldsymbol{x}$, where $\mathrm{Tr}(\boldsymbol{P}_{s})$ is the trace of the pressure tensor for species $s$. The drift (i.e.\ bulk) energy of the generic species $s$ is defined as $E^{\mathrm{d}}_{s} \doteq (1/2) \int \rho_{s} V_{s}^2 ~\mathrm{d}^3 \boldsymbol{x}$, where $\boldsymbol{V}_{s} \doteq (1/n_{s}) \int \boldsymbol{v} f_{s} ~\mathrm{d}^3 \boldsymbol{v}$ represents the drift velocity of the $s$-th species and $f_{s}$ denotes the distribution function of the $s$-th species. The pressure tensor is given by $\boldsymbol{P}_{s} \doteq m_{s} \int (\boldsymbol{v} - \boldsymbol{V}_{s})(\boldsymbol{v} - \boldsymbol{V}_{s}) f_{s} ~\mathrm{d}^3 \boldsymbol{v}$, where $\boldsymbol{v}$ denotes the particle velocity. In a reference system aligned with the direction $\hat{\boldsymbol{b}} \doteq \boldsymbol{B} /\lVert \boldsymbol{B} \rVert$ of the magnetic field, the perpendicular pressure is thus defined as $p_{\perp, s} \doteq \boldsymbol{P}_{s} : (\mathbb{1} - \hat{\boldsymbol{b}} \hat{\boldsymbol{b}})/2$, and the parallel pressure is $p_{\parallel, s} \doteq \boldsymbol{P}_{s} : \hat{\boldsymbol{b}} \hat{\boldsymbol{b}}$. The thermal energy for the generic species $s$, in the magnetic field-aligned reference frame, is calculated in the perpendicular direction as $E^{\mathrm{th}}_{\perp, s} \doteq (1/(2 k_{\mathrm{B}})) \int p_{\perp, s} \, \mathrm{d}^3 \boldsymbol{x}$ and in the parallel direction as $E^{\mathrm{th}}_{\parallel, s} \doteq (1/(2 k_{\mathrm{B}})) \int p_{\parallel, s} \, \mathrm{d}^3 \boldsymbol{x}$. The components of kinetic energy in Figure~\ref{fig:energy}(b) are: the beam parallel and perpendicular thermal energy, respectively $\delta E^{\mathrm{th}}_{\parallel, \mathrm{b}}/U_{0}$ and $\delta E^{\mathrm{th}}_{\perp, \mathrm{b}}/U_{0}$, core parallel and perpendicular thermal energy, respectively $\delta E^{\mathrm{th}}_{\parallel, \mathrm{c}}/U_{0}$ and $\delta E^{\mathrm{th}}_{\perp, \mathrm{c}}/U_{0}$, and the drift energy of both the core and the beam $\delta E^{\mathrm{d}}_{\mathrm{c}}/U_{0}$ and $\delta E^{\mathrm{d}}_{\mathrm{b}}/U_{0}$. 

During the \textit{non-linear phase}, the magnetic energy is converted back into kinetic energy, as illustrated in Figure~\ref{fig:energy}(a). As typically observed after instability saturation, the driver behind the instability onset is removed by the instability itself. In Figure~\ref{fig:energy}(b), it is evident that $\delta E^{\mathrm{th}}_{\perp, \mathrm{c}}/U_{0}$ and $\delta E^{\mathrm{th}}_{\parallel, \mathrm{c}}/U_{0}$ do not change significantly. Meanwhile, $\delta E^{\mathrm{th}}_{\perp, \mathrm{b}}/U_{0}$ is steadily increasing, while $\delta E^{\mathrm{d}}_{\mathrm{b}}/U_{0}$ is consistently decreasing. During this phase, the bulk energy stored in the beam's drift motion is converted into thermal perpendicular energy of the beam. A key parameter in proton beam instabilities is the core-to-beam drift velocity, which determines the growth rate of the instability and the types of waves produced \citep{shaabanElectromagneticIonIon2020}. The electron population, as shown in Figure~\ref{fig:energy}(c), does not significantly contribute to the system's dynamics. The electron bulk energy remains nearly constant around zero, with a slight increase during the \textit{quasi-linear} phase. The thermal energy, instead, monotonically increases throughout the entire time of the simulation. However, the magnitudes of electron drift and thermal energy are notably smaller compared to those of the protons (both core and beam). Therefore, we conclude that there is no interplay between protons and electrons in our case.

In Figure~\ref{fig:energy}(d)-(f), we present the time evolution of the proton temperature anisotropy for core, beam and the total proton populations. In the latter case, we consider core plus beam as a single proton population, and we calculate the moments of this single population rather than the moments for core and beam separately. The perpendicular and parallel pressures for the total population of protons are calculated from the pressure tensor $\boldsymbol{P}_{\mathrm{p}} \doteq m_{\mathrm{p}} \int (\boldsymbol{v} - \boldsymbol{V}_{\mathrm{cm}, \mathrm{p}})(\boldsymbol{v} - \boldsymbol{V}_{\mathrm{cm}, \mathrm{p}}) f_{\mathrm{p}} \, \mathrm{d}^3 \boldsymbol{v}$, where the velocity of the center of mass of the proton population is $\boldsymbol{V}_{\mathrm{cm}, \mathrm{p}} \doteq \left(n_{\mathrm{c}}\boldsymbol{V}_{\mathrm{c}}+n_{\mathrm{b}}\boldsymbol{V}_{\mathrm{b}}\right)/\left(n_{\mathrm{c}}+n_{\mathrm{b}}\right)$. From these results, shown in panels (d)-(f), we infer that the pronounced outcome of the considered instability is the increase in proton temperature anisotropy in the \textit{quasi-linear phase}, for both the core (Figure~\ref{fig:energy}(d)) and the beam (Figure~\ref{fig:energy}(d)) proton populations, and consequently this is reflecting on the total proton population (Figure~\ref{fig:energy}(f)). Therefore, the energy is redistributed from the parallel to the perpendicular direction. 

In conclusion, studying the energy exchanges of the system in Figure~\ref{fig:energy}(a)-(f) reveals how energy is redistributed inside the system. The two primary sources of free energy are the beam temperature anisotropy and the relative drift speed of the core and beam. Therefore, based on this analysis, we can conclude that the system evolves through a proton-beam instability, also known as ion-ion instability, which is powered not only by thermal anisotropy but also by the drifting population, resembling a beam-plasma type of instability.

\begin{figure*}
\includegraphics[width=1\textwidth]{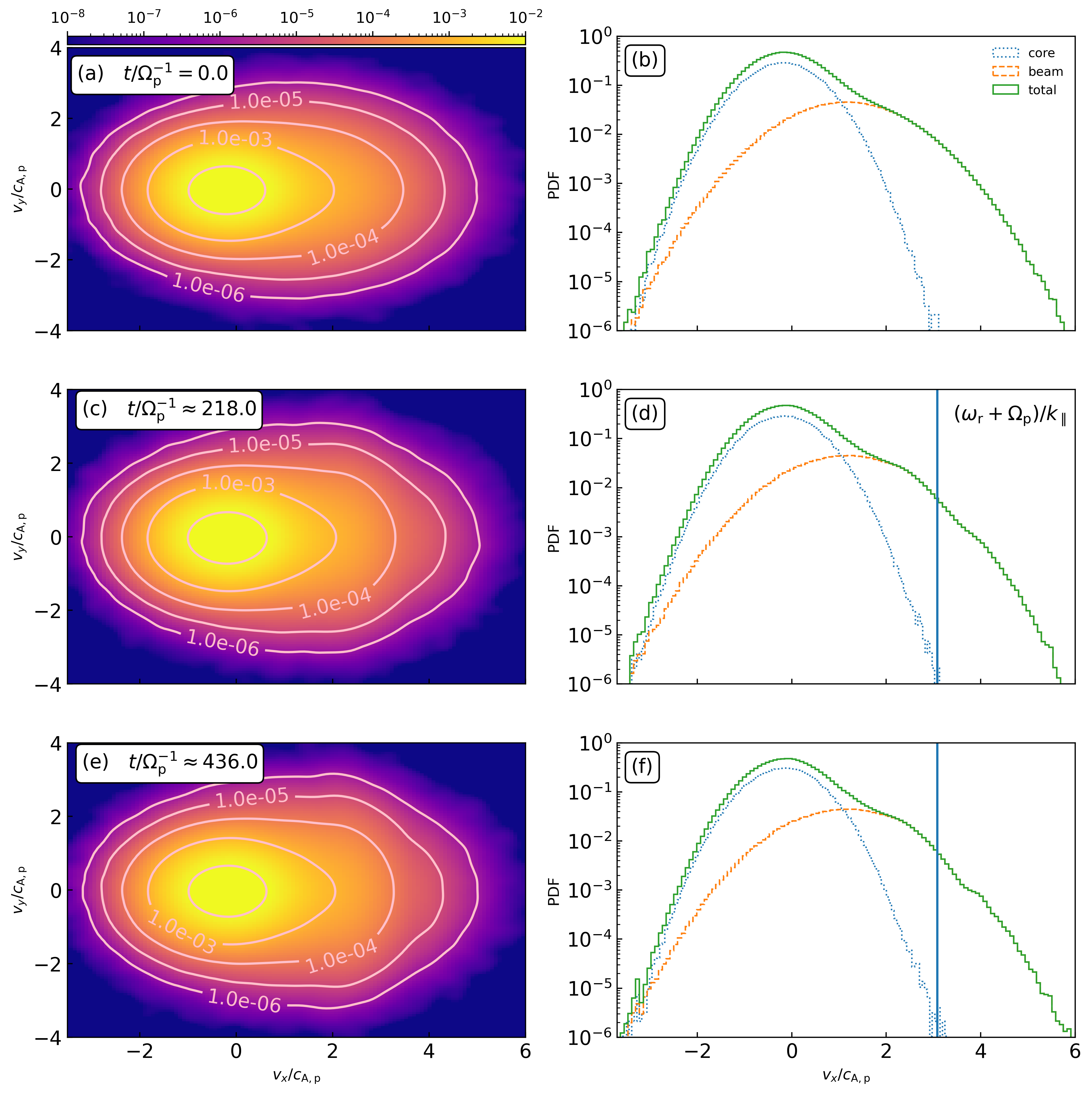}
\caption{Evolution of the proton VDF $f_{\mathrm{p}}(v_{x},v_{y},t)$ (left column) and its parallel cut $f_{\mathrm{p}}(v_{x},v_{y}=0,t)$ (right column). Panels (a)--(b) represent the \textit{quasi-stationary phase} of the proton-beam instability instability. Panel (a) shows the initial distribution of protons $f_{\mathrm{p}}(v_{x}, v_{y}, t)$, whereas panel (b) shows the horizontal cut $f_{\mathrm{p}}(v_{x}, v_{y}=0, t)$ (with the core and beam sub-populations indicated by different lines), at $t/\Omega_{\mathrm{p}}^{-1}= 0$. Panels (c)--(d) represent the phase space of protons during the \textit{quasi-linear phase} at $t/\Omega_{\mathrm{p}}^{-1} \approx 218$. The proton VDF during the \textit{non-linear phase} at $t/\Omega_{\mathrm{p}}^{-1} \approx 436$ is shown in panels (e) and (f). The resonant velocity of the system, $v_{\mathrm{res}}=(\omega_{\mathrm{r}}+\Omega_{\mathrm{p}})/k_{\parallel}$, is shown by the blue line in panel (f). \label{fig:vdf}}
\end{figure*}

Figure~\ref{fig:vdf} shows the time evolution of the core, beam, and total proton VDF, denoted as $f_{\mathrm{p}}(v_{x}, v_{y}, t)$, in the left column, as well as a one-dimensional cut $f_{\mathrm{p}}(v_{x},v_{y}=0,t)$, in the right column, to provide further evidence of the wave-particle resonant interaction. Panels (a) and (b) in Figure~\ref{fig:vdf} represent the initial conditions of the simulation, corresponding to the beginning of the \textit{quasi-stationary phase} of the instability. Subsequently, panels (c) and (d) in Figure~\ref{fig:vdf} at $t/\Omega_{\mathrm{p}}^{-1} \approx 218$ illustrate the \textit{quasi-linear phase}, during which the beam population undergoes resonant interaction with the excited fast magnetosonic waves. Figure~\ref{fig:vdf}(e)-(f) represents the proton VDF during the \textit{non-linear phase} at $t/\Omega_{\mathrm{p}}^{-1} \approx 436$. A hint of the anisotropic proton beam (i.e.\ extensions of the beam towards the values of higher $v_y$), initially observed forming during the \textit{quasi-linear phase} in Figure~\ref{fig:vdf}(c) and (d), becomes more pronounced during the \textit{non-linear phase} in Figure~\ref{fig:vdf}(e)--(f). In Figure~\ref{fig:vdf}(d) and (f), we have shown a vertical line indicating the resonant velocity $v_{\mathrm{res}} \doteq (\omega_{\mathrm{r}} - o\Omega_{\mathrm{p}})/k_{\parallel}$, where $o\in\mathbb{N}$, for the fast-magnetosonic wave quantities: $k_{\parallel}/d_{\mathrm{p}^{-1}} \approx 0.501$, $\Omega_{\mathrm{p}}/\omega_{\mathrm{p}} \approx 0.0029$, $\omega_{\mathrm{r}}/\omega_{\mathrm{p}} \approx 0.0015$. Cyclotron resonance with ${o}=-1$, which corresponds to the backward-propagating Alfv\'en wave and the forward-propagating fast-magnetosonic mode \citep{kleinInferredLinearStability2021}, has been assumed. Comparing this to the distribution in Figure~\ref{fig:vdf}(b), we observe a decrease in the number of particles for velocities $v > v_{\mathrm{res}}$ which identifies a phase space depletion region. Conversely, we observe an increase in the number of particles for velocities $v < v_{\mathrm{res}}$ creating a phase space pile-up region.

\begin{figure}
\includegraphics[width=1\columnwidth]{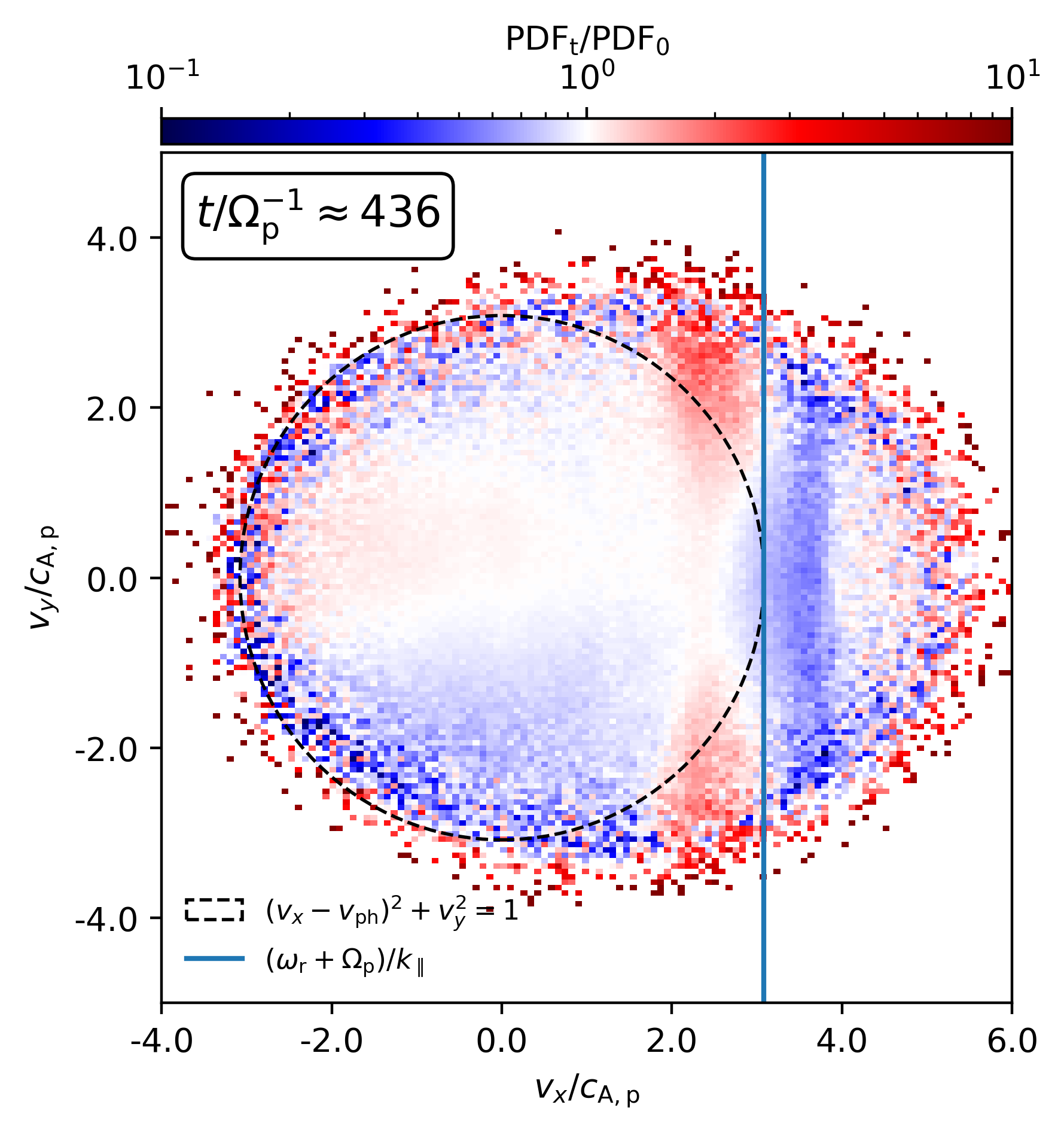}
\caption{Proton VDF $f_{\mathrm{p}}(v_{x}, v_{y}, t)$ at time $t/\Omega_{\mathrm{p}}^{-1} \approx 436$ normalized by its value at the initial time. The blue line shows the parallel resonant velocity $v_\mathrm{r}$. The black dashed circle centered at $(v_{x}, v_{y})=(v_{\mathrm{ph}}, 0)$, with $v_{\mathrm{ph}}=\omega_{\mathrm{r}}/k_{\parallel}$ the parallel phase velocity of the waves, is the diffusion path due to the resonant interaction. \label{fig:vdfn}}
\end{figure}

Here, we will investigate the nature of proton diffusion through wave-particle resonant interaction. In Figure~\ref{fig:vdfn}, the proton VDF is shown at time $t/\Omega_{\mathrm{p}}^{-1}= 436$, normalized by its value at the initial time. The normalization allows us to highlight small changes in the VDF more easily. In the white regions, no significant particle diffusion occurs, while in the blue regions, there are fewer particles than in the initial state, and the red regions are where particles have accumulated. The blue vertical line here indicates the parallel resonant velocity, while the black circle, $(v_{x}-v_{\mathrm{ph}})^2+v_{y}^2=1$, represents an iso-line defining the path in velocity phase space where particle energy remains constant. Here, $v_{\mathrm{ph}}=\omega_{\mathrm{r}}/k_{\parallel}$ denotes the parallel phase velocity of the waves. Particles diffuse in phase space along trajectories tangent to the iso-line of constant energy. Resonant particles experience a gain in kinetic energy, manifested as an increase in $v_{x}^2+v_{y}^2$. This acquired energy is extracted from the waves at resonant frequency $\omega_\mathrm{res}$, resulting in their damping. As illustrated in Figure~\ref{fig:vdfn}, a VDF depletion region (blue) is observed for velocities $v > v_{\mathrm{res}}$, while a phase space pile-up region occurs for $v<v_{\mathrm{res}}$.

\section{Discussion and Conclusions}\label{sec:discussion}

We presented the first fully kinetic study of unstable proton VDFs from the recent PSP observation of \nth{30} January 2020. We employed the \textsc{ECsim} code, a semi-implicit exactly energy conserving PIC  \citep{lapentaExactlyEnergyConserving2017,gonzalez-herreroECsimCYLEnergyConserving2019,bacchiniRelSIMRelativisticSemiimplicit2023,croonenExactlyEnergyconservingElectromagnetic2024} in order to confirm and extend to the non-linear regime the previous analysis by \citealt{kleinInferredLinearStability2021}, studying in detail the energy exchanges between the different proton species. 
For simplicity, we initialised the electrons as a Maxwellian distribution, notwithstanding the very non-thermal nature of the electron VDF in the solar wind \citep{miceraParticleincellSimulationWhistler2020,miceraRoleSolarWind2021, verscharenElectronDrivenInstabilitiesSolar2022}.
In reality, electrons can exhibit more complex structures than a simple Maxwellian distribution, such as temperature anisotropies, core/Strahl configurations, or Kappa distributions. All those possibilities could modify the evolution of proton instabilities (e.g.\ \citealt{shaabanElectromagneticIonIon2020,miceraParticleincellSimulationsParallel2020,lopezMixingSolarWind2022}). In this work, we chose a simple Maxwellian to isolate the proton evolution. However, the proton evolution could still affect the electron dynamics. The proton population was modeled using a double-component bi-Maxwellian distribution, as described in \citealt{kleinInferredLinearStability2021}. In this representation the anisotropic proton core and proton beam species drift along a background magnetic field. Proton beams are known to drive electromagnetic proton-beam instabilities, leading to the generation of waves (e.g.\ \citealt{garyElectromagneticIonBeam1984}). In this case, the system develops a RH resonant ion–ion instability with maximum growth rate $\gamma_{\mathrm{m}}/ \omega_{\mathrm{p}} = 0.000121$ and wavenumbers $k_{\parallel}/d_{\mathrm{p}}^{-1}=0.501$ and $k_{\perp}/d_{\mathrm{p}}^{-1}=0.0245$. This corresponds to a fast-magnetosonic wave which is RH circularly polarised and compressible, in good agreement with the results found by \citealt{kleinInferredLinearStability2021}.

Based on our analysis of the temporal evolution of the energy, we can summarise a few key points. The proton core sub-population does not contribute significantly to the dynamics of the system, which is primarily dominated by the beam sub-population. The thermal energy of the beam protons is transferred from the parallel to the perpendicular direction. The parallel thermal energy of the proton beam increases at the expense of the proton beam drift energy, which constitutes the main source of free energy. During the non-linear phase, the beam drift energy is contributing to the increase in perpendicular thermal energy of the beam. The notable result of the wave-particle interaction is an increase in proton temperature anisotropy, redistributing energy from the parallel to the perpendicular direction.

In our simulated proton VDFs, starting from the quasi-linear phase, we note a reduction in the number of particles for velocities exceeding the resonant velocity $v_{\mathrm{res}}$, indicating a phase space depletion region. Conversely, there is an increase in the particle count for velocities below $v_{\mathrm{res}}$, manifesting as a phase space pile-up region. Our simulation indicates that the fast magnetosonic waves are inducing resonance in the proton beam population. Consequently, these protons experience diffusion in phase space, leading to perpendicular heating effects as they move into less densely populated regions, resulting in larger perpendicular velocity \citep{verscharenSelfinducedScatteringStrahl2019}. The diffusion process initiates in the quasi-linear phase and continues into the non-linear phase, where it saturates, gradually, diminishing in intensity.

Cyclotron resonance plays a significant role in shaping magnetic field spectra and distribution functions observed in the solar wind \citep{goldsteinPropertiesFluctuatingMagnetic1994, leamonContributionCyclotronresonantDamping1998}. The nonlinear evolution of the fast magnetosonic mode is significantly influenced by cyclotron resonance, which plays a crucial role in inducing perpendicular proton heating and redistributing energy from the parallel to the perpendicular direction. This resonance process effectively mitigates the free energy associated with the parallel anisotropy, constituting a new channel of regulation of the solar wind temperature anisotropy. We demonstrated that wave-particle resonant interaction can modify the proton beam VDF, inducing perpendicular heating that reaches up to 12-15\% of its original value. According to \citealt{vernieroStrongPerpendicularVelocityspace2022}, the ``hammerhead'' features in the observed VDFs are the result of proton beam driven instabilities of the parallel FM/W instability. The strongly asymmetric proton beam exhibits a temperature anisotropy in the range of 2 to 8. We speculate that our simulation could produce a more pronounced proton beam anisotropy by considering the presence of turbulence, which is ubiquitous in the solar wind. This turbulence could provide an additional energy source, dissipated through interaction with the proton beam, thereby enhancing the anisotropy. It is known that turbulence drives ion heating \citep{servidioLocalKineticEffects2012,matthaeusPathwaysDissipationWeakly2020,arroSpectralPropertiesEnergy2022}, inducing anisotropic deformations on Maxwellian particle distributions \citep{servidioKineticModelPlasma2015,arroGenerationSubionScale2023,arroLargescaleLinearMagnetic2024}. However, the interplay between turbulence and non-Maxwellian distributions, like those studied in the present work, has been poorly investigated and could significantly impact dissipation. This point will be addressed in future works.

The work by \citealt{ofmanModelingIonBeams2022} presents a comparison between 2D and 3D hybrid simulations of ion beam instability in solar wind plasma, based on observations from the PSP near perihelia. That study demonstrated that 3D simulations do not exhibit additional effects compared to 2D simulations, indicating that three-dimensional effects do not significantly affect proton-beam-driven instabilities. However, solar wind protons may be more accurately described by a kappa distribution function instead of a Maxwellian distribution \citep{pierrardKappaDistributionsTheory2010, livadiotisUnderstandingKappaDistributions2013, nicolaouDeterminationKappaDistribution2020}. A natural progression of this work could involve extending the simulation with a study of this system modeling the proton beam using a kappa distribution. However, these aspects are beyond the scope of this paper and are left for future studies.

\section*{Acknowledgements}\label{sec:acknow}
During the preparation of this work, we were deeply saddened by the passing of our colleague and friend, Gianni Lapenta. We dedicate this paper to his memory.

The computational resources and services used in this work were partially provided by the VSC (Flemish Supercomputer Center), funded by the Research Foundation Flanders (FWO) and the Flemish Government – department EWI.
L.P.\ acknowledges support from a PhD grant awarded by the Royal Observatory of Belgium and the Fonds voor Wetenschappelijk Onderzoek (FWO), with id number 11PCB24N.
A.N.Z.\ thanks the Belgian Federal Science Policy Office (BELSPO) for the provision of financial support in the framework of the PRODEX Programme of the European Space Agency (ESA) under contract number 4000136424.
F.B.\ acknowledges support from the FED-tWIN programme (profile Prf-2020-004, project ``ENERGY'') issued by BELSPO, and from the FWO Junior Research Project G020224N granted by the Research Foundation -- Flanders (FWO).
A.M. is supported by the Deutsche Forschungsgemeinschaft (German Science Foundation; DFG) project 497938371.
M.E.I.\ acknowledges support from DFG within the Collaborative Research Center SFB1491. 

\newpage
\bibliography{References}{}
\bibliographystyle{aasjournal}
\end{document}